\documentclass[aps,prd,
preprintnumbers,nofootinbib,superscriptaddress]{revtex4}
\usepackage[dvips]{graphicx}
\usepackage{caption}
\usepackage{ragged2e}
\usepackage{subcaption}
\usepackage{bm,latexsym,amsmath,amssymb,amsfonts,color}
\usepackage{ragged2e}
\begin{document}

\title{Exact charged and rotating toroidal black hole in the \\ Einstein $SU(N)$-Skyrme model}

\author{Carla Henr\'iquez-Baez}
\email{carla.henriquez@umayor.cl}
\affiliation{Centro Multidisciplinario de F\'isica, Vicerrector\'ia de Investigaci\'on, Universidad Mayor, Camino La Pir\'amide 5750, Santiago, Chile.}
\author{Marcela Lagos}
\email{marcela.lagos@uss.cl}
\affiliation{Universidad San Sebastián, Avenida del Cóndor 720, Santiago, Chile.}
\author{Evelyn Rodr\'iguez}
\email{erodriguez@ucsc.cl}
\affiliation{Departamento de Matem\'atica y F\'isica Aplicadas,
Universidad Cat\'olica de la Sant\'isima Concepci\'on,
Alonso de Ribera 2850, Concepci\'on, Chile}
\affiliation{Grupo de Investigación en Física Teórica, GIFT, Universidad Católica de la Santísima Concepción, Alonso de Ribera 2850, Concepción, Chile.}
\author{Aldo Vera}
\email{aldo.vera@umayor.cl}
\affiliation{N\'ucleo Matem\'aticas F\'isica y Estad\'istica, Universidad Mayor, Avenida Manuel Montt 367, Santiago, Chile}
\affiliation{Centro Multidisciplinario de F\'isica, Vicerrector\'ia de Investigaci\'on, Universidad Mayor, Camino La Pir\'amide 5750, Santiago, Chile.}

\begin{abstract}
In this paper, we construct an exact solution of the Einstein $SU(N)$-Skyrme model in $D=4$ space-time dimensions describing a charged and rotating black hole with toroidal horizon. Rotation is added by applying an improper coordinate transformation to the known static toroidal black hole with Skyrme hair, while the electric charge is supplemented by considering a $U(1)$ gauge field interacting with Einstein gravity. We perform the thermal analysis in the grand canonical ensemble, explicitly showing the role that the flavor number plays. Some discussions about stability are also considered.
\end{abstract}

\maketitle

\newpage

%%%%%%%%%%%%%%%%%%%%%%%%%%%%%%%%%%%%%%%
\section{Introduction} \label{sec-1}
%%%%%%%%%%%%%%%%%%%%%%%%%%%%%%%%%%%%%%%

Hairy black holes are solutions of theories of gravity characterized by more parameters, in addition to their mass, charge, and angular momentum. Their \textit{hair} can be defined as a free parameter that does not come from a field of gravitational or electromagnetic nature, or alternatively, as a quantity that is not subject to a Gauss law.
These kinds of configurations can be constructed by sorting some of the assumptions of the well-known no hair conjecture \cite{Ruffini:1971bza}, \cite{Bekenstein:1996pn}. By means of analytical or numerical methods, hairy black holes have been found by coupling general relativity (GR) or alternative theories of gravity with appropriate matter fields, introducing a cosmological constant, or considering non-minimal couplings.

The first counterexample to the no-hair conjecture was reported in Ref. \cite{Luckock:1986tr} (see also Ref. \cite{Droz:1991cx}), where a spherically symmetric hairy black hole was numerically constructed considering the coupling of GR with the $SU(2)$-Skyrme model. Soon after, other hairy black holes were constructed using non-abelian gauge fields; see \cite{Bizon1}-\cite{Torii:1993vm}, and references therein.
The black hole constructed by Luckock and Moss in Ref. \cite{Luckock:1986tr} is particularly interesting, not only because it is stable under spherical linear perturbations \cite{Heusler:1992av}, but also because the matter field considered has strong theoretical and phenomenological support.
In fact, the Skyrme model \cite{Skyrme1}, \cite{Skyrme2} is one of the most relevant effective field theories since it describes the low-energy limit of quantum chromodynamics (QCD) in the large $N_c$ expansion (see Refs. \cite{MantonBook} and \cite{WeinbergBook}). From an action constructed from a complex scalar field $U(x)\in SU(N)$, being $N$ the flavor number of QCD, it allows us to describe baryons from the non-linear interaction between mesons \cite{Witten}, \cite{ANW}.

In recent years, other gravitational configurations have been constructed in the Skyrme model, including black holes, compact stars, gravitating solitons, extended objects, among others. Relevant numerical solutions were reported in Refs. \cite{Bizon2}-\cite{Leask} (see references therein), while exact solutions have been constructed in Refs. \cite{CanforaMaeda}-\cite{toroidal4}, inspired by results derived in flat space-time describing inhomogeneous condensates of topological solitons, as can be seen in Refs. \cite{Pedro1}-\cite{Canfora:2023zmt}.

In particular, in Ref. \cite{toroidal4} an exact solution describing a hairy toroidal black hole in the $SU(N)$-Skyrme model was found (see also Refs. \cite{toroidal1}, \cite{toroidal2} and \cite{toroidal3}). Such a black hole is characterized by allowing an arbitrary number of flavors and being asymptotically locally anti-de Sitter (AdS). Relevant properties of this family of black holes have been recently explored in Refs. \cite{Abbas}-\cite{Al-Badawi}.

Black holes with toroidal horizon are solutions of theories of gravity allowed by the presence of a cosmological constant \cite{Toro1}-\cite{Anabalon:2022ksf} (see Ref. \cite{Toro8} for a review). Recently, these kinds of configuration have attracted a lot of attention due to their applications in holography (see, for instance, Refs.  \cite{Papantonopoulos:2011zz}-\cite{McInnes:2022pig}). The reason behind this is the possibility of using the AdS/CFT correspondence
to describe interesting field theories on the boundary of a black hole space-time. In particular, the quark-gluon plasma (QGP) is modeled, via holography, by a field theory dual to a thermal AdS black hole. Since the QGP exists in Minkowski space-time, the use of black holes with topologically planar event horizons is necessary. In this context, toroidal black holes coming from the Skyrme model could be of great relevance due to the non-perturbative nature of the model.

In this paper, we generalize the hairy toroidal black hole reported in Ref. \cite{toroidal4} to a charged and rotating solution. We show that even when rotation, electric charge, and an arbitrary number of flavors are considered, the solution can still be found analytically. In addition, we analyze the thermal behavior and global stability of this novel solution. Here, it is important to emphasize that the description of rotating black hole solutions in $D=4$ space-time dimensions is not only of great astrophysical interest, it also constitutes an important problem in theoretical physics that usually requires the use of numerical methods, even for relatively simple coupled matter fields.

This paper is organized as follows: In Sec. \ref{sec-2} we give a brief summary of the Einstein-Skyrme model, and introduce the Ansatz for the pionic field that will be used in the following sections. Also, we review the exact toroidal black hole solution with Skyrme hair, and present the thermal formulation to be used. In Sec. \ref{sec-3} we construct a hairy rotating toroidal black hole, and in Sec. \ref{sec-4} we construct a hairy charged toroidal black hole, performing thermal analysis in both cases. Sec. \ref{sec-5} is devoted to the construction of the general solution describing a charged and rotating toroidal black hole and to the discussion of its global stability in the grand-canonical ensemble.

%%%%%%%%%%%%%%%%%%%%%%%%%%%%%%%%%%%%%%%
\section{Preliminaries}  \label{sec-2}
%%%%%%%%%%%%%%%%%%%%%%%%%%%%%%%%%%%%%%%

\subsection{The Einstein-Skyrme model}
%%%%%%%%%%%%%%%%%%%%%%%%%%%%%%%%%%%%%%%%%%%%%%%

We consider the Einstein $SU(N)$-Skyrme gravity model in $D=4$ space-time dimensions, described by the action 
\begin{equation}
I[g,U]=\int_{\mathcal{M}} d^{4}x\sqrt{-g}\left( \frac{R-2\Lambda }{2\kappa }+%
\frac{K}{4}\mathrm{Tr}[L^{\mu }L_{\mu }]+\frac{K\lambda }{32}\mathrm{Tr}%
\left( G_{\mu \nu }G^{\mu \nu }\right) \right) \ .  \label{I}
\end{equation}
Here $R$ is the Ricci scalar, $g$ is the determinant of the metric $g_{\mu\nu}$, and  $L_{\mu}$ are the Maurer-Cartan left-invariant form components, given by 
\begin{equation}
L_{\mu}=U^{-1}\partial_{\mu} U = L^{i}_{\mu}t_{i} \ , \label{Maurer-Cartan-form}
\end{equation}
where $U\left(x \right) \in SU(N)$ is the pionic field, being $N$ the flavor number. The matrices $t_{i}$ are the generators of the $SU(N)$ Lie group, with $i=1, \ldots , \left( N^2-1 \right)$, and the $G_{\mu\nu}$ tensor is defined in terms of the left-invariant form components as 
\begin{equation}
G_{\mu \nu }=\left[ L_{\mu },L_{\nu }\right] \ .
\end{equation}
In Eq. \eqref{I}, $\kappa$ is the gravitational constant, $\Lambda$ is the cosmological constant, $K$ is defined in terms of the pion decay constant as $K=\frac{f_\pi^2}{4}$, and $\lambda$ is a positive number fixed experimentally.

The corresponding field equations of the Einstein $SU(N)$-Skyrme model are the following:
\begin{gather} \label{Eq1}
\nabla _{\mu }L^{\mu }+\frac{\lambda }{4}\nabla _{\mu }[L_{\nu },G^{\mu \nu
}]=0\ , \\  \label{Eq2}
R_{\mu \nu }-\frac{1}{2}Rg_{\mu \nu }+\Lambda g_{\mu \nu }=\kappa T_{\mu \nu
}\ ,
\end{gather}
where the energy-momentum tensor, $T_{\mu\nu}$, is given by
\begin{equation}
T_{\mu \nu }=-\frac{K}{2}\mathrm{Tr}\left[ L_{\mu }L_{\nu }-\frac{1}{2}%
g_{\mu \nu }L^{\alpha }L_{\alpha }\right. \,+\left. \frac{\lambda }{4}\left(
g^{\alpha \beta }G_{\mu \alpha }G_{\nu \beta }-\frac{g_{\mu \nu }}{4}%
G_{\sigma \rho }G^{\sigma \rho }\right) \right] \ .  \notag  \label{Tmunu}
\end{equation}

\subsection{$SU(N)$ Skyrme fields}
%%%%%%%%%%%%%%%%%%%%%%%%%%%%%%%%%%%%%%%%%%%%%%%

In order to construct exact solutions with an arbitrary number of flavors, we will use for the pionic field $U(x)$ the \textit{maximal embedding} of $SU(2)$ into $SU(N)$ in the Euler angles parametrization \cite{euler1}-\cite{euler4}. This Ansatz is particularly useful as it allows the construction of solutions where the $N$ parameter appears explicitly; this, using as basis three $N\times N$ matrices which gives rise to an irreducible spin-$j$ representation of $SU(2)$ of spin $j=(N-1)/2$. Recently, this formalism has been applied to the study of inhomogeneous condensates in the Skyrme model \cite{Pedro2}-\cite{SU(N)2}, as well as to the construction of black objects in GR supported by bosonic and fermionic matter content \cite{toroidal3}, \cite{toroidal4}, \cite{Gomberoff}.

Following Refs. \cite{euler1}-\cite{euler3}, the matter field $U(x)\in SU(N)$ in the Euler angles parametrization reads  
\begin{equation}
    U=e^{F_{1}\left(x^{\mu} \right)\cdot T_{3}}e^{F_{2}\left(x^{\mu}\right) \cdot T_{2}} e^{F_{3}\left(x^{\mu}\right) \cdot T_{3}} \ ,  \label{matter-ansatz}
\end{equation}
where $F_i$ are arbitrary functions of the space-time coordinates, and  $T_{i}$ (with $i=1,2,3$) are the generators of a $3$-dimensional sub-algebra of $\mathfrak{su}(N)$, which are explicitly given by 
\begin{align}
T_1&=-\frac{i}{2}\sum_{j=2}^{N} \sqrt{(j-1)(N-j+1)}(E_{j-1,j}+E_{j,j-1}) \ ,
\label{T1} \\
T_2&=\frac{1}{2}\sum_{j=2}^{N} \sqrt{(j-1)(N-j+1)}(E_{j-1,j}-E_{j,j-1}) \ ,
\label{T2} \\
T_3&=i\sum_{j=1}^{N} (\frac{N+1}{2}-j)E_{j,j} \ ,  \label{T3}
\end{align}
with $(E_{i,j})_{mn}=\delta_{im}\delta_{jn}$, and being $\delta_{ij}$ the Kronecker delta. The above matrices satisfy the following commutation relation
\begin{equation}
\left[T_{j}, T_{k} \right] \  = \  \epsilon_{jkm}T_{m} \ . 
 \end{equation}
One way to see that configurations constructed using this Ansatz are in fact non trivial embeddings of $SU(2)$ into $SU(N)$ is through the computation of the trace of the generators in Eqs. \eqref{T1}, \eqref{T2} and \eqref{T3}, that is
\begin{equation} \label{Tr}
\text{Tr}\left(T_{j}T_{k} \right) \  = \ - \frac{1}{2} a_N\delta_{jk} \ , \qquad a_N = \frac{N\left(N^2-1 \right)}{6} \ . 
\end{equation}
From Eq. \eqref{Tr} we see that the trace of the generators  -quantity that appears, for instance, in the energy-momentum tensor in Eq. \eqref{Tmunu}- (and which characterizes the corresponding field configurations), depends explicitly on the $N$ parameter. This means that matter fields with different values of $N$ leads to different solutions with a particular spin. The key point in this construction is that, actually, the map between the Lie groups is not an embedding of $SU(2)$ into $SU(N)$, but just of $S_3$ into $SU(N)$; see Refs. \cite{euler1}-\cite{euler3} for mathematical details. 

\subsection{The toroidal black hole} 
%%%%%%%%%%%%%%%%%%%%%%%%%%%%%%%%%%%%%%%

Recently, in Ref. \cite{toroidal4}, it was presented an exact solution of the Einstein-Skyrme model describing a toroidal black hole with Skyrme hair for arbitrary values of  $N$ in $D=4$ space-time dimensions. According to Ref. \cite{toroidal4}, the space-time is described by a static metric with toroidal horizon 
\begin{equation}  \label{g1}
 ds^2 =  -f(r) dt^2+ \frac{1}{f(r)} dr^2 + r^2 d\theta^2 + r^2 d\phi^2 \ ,
\end{equation}
where the angular coordinates $\{\theta$, $\phi\}$ have the ranges $0\leq\theta < \pi$, $0\leq \phi < 2\pi$.
In order to find an exact solution, the functions $F_{i}$ in Eq. \eqref{matter-ansatz} were chosen as 
\begin{equation} \label{matter1}
 F_1(x^\mu) = 0 \ , \qquad F_2(x^\mu) =  q \theta \ , \qquad F_3(x^\mu) = q \phi \ ,   
\end{equation}
where $q$ is an integer number.\footnote{In principle, the functions $F_2$ and $F_3$ can have different constants, lets say $q_1$ and $q_2$, and the complete equations system is still solved. However, this new hair parameter introduces a possible conical singularity in the metric through the presence of a term in the form $q_1^2/q_2^2$ in front of the torus metric. This conical singularity is removed by considering $q_1=q_2 \equiv q$, as in Eq. \eqref{matter1}.} With the Ansatz in Eqs. \eqref{g1} and \eqref{matter1}, one can check that the Skyrme equations in Eq. \eqref{Eq1} are automatically satisfied, while the Einstein equations in Eq. \eqref{Eq2} are solved analytically providing the following expression for the lapse function,
\begin{equation} \label{f1}
f(r)= -\frac{K \kappa q^2 a_N}{4}-\frac{m}{r}  +\frac{K \kappa q^4 \lambda a_N}{32 r^{2}}  - \frac{\Lambda}{3} r^2 \ .
\end{equation}
Here $m$ is an integration constant (related to the mass) and $a_N$ has been defined in Eq. \eqref{Tr}. This solution represents an asymptotically locally AdS toroidal black hole with Skyrme hair (being $q$ the hair parameter), allowing arbitrary values of the flavor number.  It corresponds to the generalization of the black hole solution reported in Ref. \cite{toroidal1} (see also Refs. \cite{toroidal2} and \cite{toroidal3}), in the context of the $SU(2)$-non-linear sigma model (NLSM). Also, in Ref. \cite{toroidal4}, the thermodynamics of this black hole and its application to the construction of black $p$-branes and self-gravitating instantons were explored. 

In the next sections, we will show that the black hole in Eq. \eqref{f1} can be generalized to rotating and electrically charged configurations. In addition, we will perform the thermal analysis using the formulas given here below. 

\subsection{Euclidean action and thermodynamic potentials} 
%%%        %                                                               %%%%%%%%%%%%%%%%%%%%%%%%%%%%%%%%%%%

To compute the temperature of the black hole, we will use the Hawking temperature formula \cite{Gibbons:1977mu}, \cite{Brown:1992br} in terms of the surface gravity $\kappa_s$, namely
\begin{equation} \label{T}
T=\frac{\kappa_s}{2\pi}=\frac{1}{2\pi}\left.\sqrt{-\frac{1}{2}\nabla_{\mu}\xi_{\nu}\nabla^{\mu}\xi^{\nu}}\right|_{r_{+}} \ , 
\end{equation}
with $\xi^{\mu}$ the killing vector.

For the Einstein-Skyrme system, the regularized Euclidean action is given by the bulk action supplemented with the Gibbons-Hawking term and the counterterms 
\begin{equation}
    I^{\text{E}}=I_{\text{bulk}}+I_{\text{GH}}+I_{\text{ct}} \ , \label{euact}
\end{equation}
where
\begin{equation}
    I_{\text{bulk}}=-\int_{\mathcal{M}} d^{4}x\,\sqrt{g}\left( \frac{R-2\Lambda }{2\kappa }+
\frac{K}{4}\mathrm{Tr}[L^{\mu }L_{\mu }]+\frac{K\lambda }{32}\mathrm{Tr}
\left( G_{\mu \nu }G^{\mu \nu }\right) \right)\,,
\end{equation}
\begin{equation}
    I_{\text{GH}}=-\frac{1}{\kappa}\int_{\partial\mathcal{M}}d^{3}x\,\sqrt{h}K_{\rho}\,, 
\end{equation}
\begin{equation}
    I_{\text{ct}}=\frac{1}{\kappa}\int_{\partial\mathcal{M}
}d^{3}x\,\sqrt{h}\left(\frac{2}{ \ell}+\frac{\ell}{2}\mathcal{R}\right)+\frac{K}{4}\int_{\partial\mathcal{M}
    }d^{3}x\,\sqrt{h}\, \ell\, \mathrm{Tr}[L^{i }L_{i}]\,.
\end{equation}
Here $h_{ij}$ is the induced metric on the boundary $\partial\mathcal{M}$ at the cut-off $r=\rho$, $\mathcal{R}$ is the Ricci scalar for the induced metric (which, in the present case, is zero) and $K_{\rho}$ is the trace of the extrinsic curvature of the boundary as embedded in $\mathcal{M}$. In addition to the well-known AdS gravitational surface terms in $I_{\text{ct}}$, we add an appropriate generalized counterterm action, according to the matter content, allowing us to end up with a regularized action; see Refs. \cite{toroidal4} and \cite{Daniel}. In addition, we have defined $\ell^{2}=-3/\Lambda$. 

Let us now move to the definition of the thermodynamic potentials from the Euclidean action. In the grand canonical ensemble, the thermodynamic quantities satisfy
 \begin{equation}  \label{Gdef}
    G(T,\Omega,\Phi)=E-TS-\Omega J-\Phi \Tilde{Q}\,, 
 \end{equation}
where $E$ is the energy, $S$ is the entropy, $J$ is the angular momentum, $\Omega$ is the angular velocity, $\Tilde{Q}$ is the electric charge and $\Phi$ is the Coulomb electric potential. The free energy of the system is computed as
\begin{equation}
    G(T,\Omega,\Phi)=\beta^{-1}I^{E}  \qquad \text{with}  \quad \beta = \frac{1}{T}  \ .
\end{equation}
Then, the extensive quantities are found to be
\begin{equation}
    S=-\left(\frac{\partial G}{\partial T}\right)_{\Omega\Phi}\,, \qquad J=-\left(\frac{\partial G}{\partial \Omega}\right)_{T \Phi}\,, \qquad \Tilde{Q}=-\left(\frac{\partial G}{\partial \Phi}\right)_{T\Omega}\,.
\end{equation}
As usual, the Coulomb electric potential $\Phi$ at the horizon is defined as
\begin{equation}
\Phi=A_{\mu}\xi^{\mu}|_{r\rightarrow\infty}-A_{\mu}\xi^{\mu}|_{r=r_+}\,,\label{elecpot}
\end{equation}
where $\xi^{\mu}$ is the Killing vector. Then, the first law of thermodynamic reads
\begin{equation}
    \delta M=T\delta S+\Omega \delta J+\Phi\delta \Tilde{Q}\,.
\end{equation}

%%%%%%%%%%%%%%%%%%%%%%%%%%%%%%%%%%%%%%%
\section{Rotating black hole} \label{sec-3}
%%%%%%%%%%%%%%%%%%%%%%%%%%%%%%%%%%%%%%%

\subsection{Adding rotation}
%%%%%%%%%%%%%%%%%%%%%%%%%%%%%%%%%%%%%%%%%%%%%%%

It is well known that it is possible to add angular momentum to static toroidal black holes by applying an \textit{improper gauge transformation} in a plane conformed by the temporal coordinate and a particular angular coordinate; see Refs. \cite{Stachel:1981fg} and \cite{MacCallum:1997ds}. Here we show that this formalism can be applied to the toroidal black hole in Eq. \eqref{f1} to get a rotating solution. However, in the present case, there is a subtlety regarding the matter field, which makes our construction different from other rotating black holes in the literature (see, for instance, Refs. \cite{rot1} and \cite{rot2}). In the present case, the Lorentz boost has to be applied also at the level of the matter field since it depends explicitly on the relevant coordinates, as can be seen from Eq. \eqref{matter1}. 

According to the above, let us consider the following improper gauge transformation in the plane $(t,\phi)$;
\begin{equation}
 t \rightarrow \frac{1}{\sqrt{1-\omega^2}}(t - \ell \omega \phi) \ , \qquad \phi \rightarrow \frac{1}{\sqrt{1-\omega^2}}\left(\phi - \frac{\omega}{\ell} t \right)\ , \label{gaugetrans}
\end{equation}
where $\omega$ is the rotation parameter (with $\omega^{2}<1$ in order to preserve the signature of the metric). Applying this transformation to the static metric in Eq. \eqref{g1}, we obtain the stationary axi-symmetric line element
\begin{equation} \label{stationary2}
ds^2 = -\frac{\left( \ell^2 f(r) - r^2 \omega^2\right)}{\ell^2 (1-\omega^2)}dt^2 + \frac{dr^2}{f(r)} + \frac{2 \omega \left(\ell^2 f(r)  - r^2 \right)}{\ell(1-\omega^2)} dt d\phi  + \frac{\left( r^2-\omega^2 \ell^2 f(r) \right)}{ (1-\omega^2)} d\phi^2 + r^2 d\theta^2\ ,
\end{equation}
where the function $f(r)$ has been defined in Eq. \eqref{f1}.
As we have mentioned before, the previous transformation also has an effect on the matter field through the function $F_{3}$, namely
\begin{equation} \label{matter2}
 F_1(x^\mu) = 0 \ , \qquad F_2(x^\mu) =  q \theta \ , \qquad F_3(x^\mu) =  \frac{q}{\sqrt{1-\omega^2}}\left(\phi - \frac{\omega}{\ell} t \right) \, .  
\end{equation}
For the Ansatz in Eqs. \eqref{stationary2} and \eqref{matter2}, one can check that the Einstein-Skyrme system is completely solved. The solution in Eq. \eqref{stationary2} represents a rotating hairy toroidal black hole, with the angular momentum given by Eq. \eqref{angmom} here below. The existence of a non-zero angular momentum makes clear that the solutions in Eqs. \eqref{g1} and \eqref{stationary2} are globally different, as expected. Indeed, it was shown in Ref. \cite{Stachel:1981fg} that the above transformation is not a permitted global coordinate transformation, instead Eq. \eqref{gaugetrans} preserves only the local geometry. Thus, the resulting space-time metric after the change in Eq. \eqref{gaugetrans}, is globally stationary but locally static. Even more, the mass of the present rotating black hole is bounded from below by the angular momentum, as can be seen from Eq. \eqref{angmom2}.

In general, the angular velocity of the zero-angular-momentum-observer (ZAMO) at any radius $r$ is determined by the quotient
$ \Omega_r = - \frac{g_{t\phi}}{g_{\phi \phi}}$.
When approaching the black hole horizon, the angular velocity turns out to be $\Omega_{H}=\frac{\omega}{\ell}$, while at spatial infinity it is $\Omega_{\infty}=0$. Then, the angular velocity entering the first law is given by 
\begin{equation} 
    \Omega=\Omega_{H}-\Omega_{\infty}=\frac{\omega}{\ell}\,.\label{angvel}
\end{equation}
We will use Eq. \eqref{angvel} for the thermodynamical analysis of our solution.

On the other hand, it is important to note that the metric can be put in the canonical form, 
\begin{equation} \label{canonical}
 ds^2 = -N^2(r) f(r) dt^2 + \frac{dr^2}{f(r)} + H(r) (d\phi+N^{\phi}dt)^2 +r^2 d\theta^2 \ ,  
\end{equation}
where 
\begin{equation}
N^2(r)= \frac{r^2(1-\omega^{2})}{r^2-\ell^2\omega^{2}f(r) } \ , \qquad N^{\phi}= -\frac{r^2-\ell^2f(r)}{r^2-\ell^2\omega^{2}f(r)}\frac{\omega}{\ell} \ ,  \qquad H(r)= \frac{r^2-\ell^2\omega^{2}f(r)}{1-\omega^{2}} \  . 
\end{equation}
One can check that all the above functions are monotonically increasing functions, and the event horizon is still determined by the roots of the function $f(r)$ (see Ref. \cite{toroidal2}). Regarding the behavior of the metric in the asymptotic region ($r\rightarrow\infty$), we see that
\begin{gather*}
    N^2(r) =1+\mathcal{O}(r^{-2})\,, \\
    N^{\phi}(r) =-\frac{K\kappa a_N \ell q^2\omega}{(1-\omega^2)r^2}+\mathcal{O}(r^{-3})\,,  \\
H(r)=r^2+\frac{K\kappa a_{N}q^2 \omega^2}{4(1-\omega^2)}+\mathcal{O}(r^{-2})\,.
\end{gather*}
Furthermore, the Ricci scalar approaches to 
\begin{equation}
    R\rightarrow -\frac{12}{\ell^2}+\mathcal{O}(r^{-2})\,.
\end{equation}
From the above we can see that the rotating hairy black hole is asymptotically locally AdS.

\subsection{Thermodynamics} 
%%%%%%%%%%%%%%%%%%%%%%%%%%%%%%%%%%%

In this subsection, following Ref. \cite{Gibbons:1977mu}, we present the thermal analysis of the rotating toroidal black hole presented above using the Euclidean approach. The first step in this formalism is to perform the following replace, in the temporal coordinate and in the rotation parameter, into Eq. \eqref{stationary2};
\begin{equation}
t\rightarrow i\tau \ , \qquad  \omega\rightarrow i\omega \ , 
\end{equation}
obtaining the corresponding Euclidean rotating black hole metric. The regularity on the horizon requires that the Wick rotated coordinate $\tau$ be identified with the period $\beta$, being $\beta$ the inverse of the temperature. Then, from Eq. \eqref{T}, the Hawking temperature of the black hole is given by
\begin{equation}\label{Temp1}
    T=-\left(\frac{K\kappa a_{N}q^{2}}{16 \pi r_{+} }+\frac{K\kappa a_{N} \lambda q^{4}}{128 \pi r_{+}^{3}}-\frac{3 r_{+}}{4 \pi\ell^2}\right)\sqrt{1-\omega^{2}} \ ,
\end{equation}
where $r_{+}$ is the largest root of the equation $\bar{f}\left(r_{+}\right)=0$ (being $\bar{f}$ the lapse function in Eq. \eqref{stationary2}), and we have used the killing vector $\xi=\partial_{t}+\Omega_{H}\partial_{\phi}$.

Taking the limit $\rho\rightarrow+\infty$, we find that the regularized Euclidean action is
\begin{equation} \label{Euclact}
    I^{\text{E}}=-\frac{\beta \pi^2 r_{+}}{4\kappa}\left(K\kappa a_N q^{2}+\frac{4 r_{+}^{2}}{\ell^2} -\frac{3K\kappa a_N \lambda q^{4}}{8 r_{+}^{2}}\right)\,,
\end{equation}
so that the free energy in Eq. \eqref{Gdef} reads
\begin{equation}\label{free}
    G=-\frac{ \pi^2 r_{+}}{4\kappa}\left(K\kappa\alpha_{N}q^{2}+\frac{4 r_{+}^{2}}{\ell^2}-\frac{3K\kappa \alpha_{N}\lambda q^{4}}{8 r_{+}^{2}}\right)\,.
\end{equation}
Then, is possible to show that the mass and angular momentum of the rotating hairy black hole are given by:
\begin{align}
E&=\frac{\pi^{2}r_+}{32\kappa \left(1-\omega^2\right)}\left[ -8 K\kappa a_N q^{2}(2-\omega^2)+\frac{32 r_{+}^{2}}{\ell^2}\left(2+\omega^2\right)+\frac{ K\kappa a_N \lambda q^{4}}{ 8 r_{+}^2}\left(2-3\omega^2\right)\right]\, ,\label{mass}\\ 
J&=-\frac{\ell\pi^2 r_{+}}{4\kappa}\left(K\kappa a_N q^{2}-\frac{12 r_{+}^{2}}{\ell^2}+
\frac{ K\kappa a_N \lambda q^{4}}{ 8 r_{+}^2}\right)\frac{\omega}{1-\omega^2}\,.\label{angmom}
\end{align}
In terms of the $m$ constant, these can be written as follows
\begin{align}
E&=\frac{\pi^{2}(2+\omega^2)}{ \kappa(1-\omega^2)}m+\frac{K \pi^2 q^2(4r_{+}^2-q^2\lambda)\omega^2}{8r_{+}(1-\omega^2)}\, ,\label{mass2}\\ 
J&=\frac{3\ell\pi^2\omega}{\kappa(1-\omega^2)}m+\frac{K \ell \pi^2 q^2(4r_{+}^2-q^2\lambda)\omega}{8r_{+}(1-\omega^2)}\,.\label{angmom2}
\end{align}
In the present case, in the absence of the matter field ($q=0$), the black hole mass does not vanish, unlike the hairy black hole reported in Ref. \cite{rot1}. Thus, in the current case, we recover the thermodynamic quantities of the rotating black hole presented in Ref.  \cite{Toro1} when $q=0$. 

On the other hand, the entropy is given by
\begin{equation}
S=\frac{\pi^{2}r_{+}^{2}}{2\sqrt{1-\omega^{2}}}=\frac{A}{4}\,.
\end{equation}
As expected, these quantities satisfy the first law of thermodynamics. In addition, we can see that the thermodynamic variables for the static solution presented in Ref. \cite{toroidal4} are recovered when $\omega=0$.

%%%%%%%%%%%%%%%%%%%%%%%%%%%%%%%%%%%
\section{Charged black hole} \label{sec-4}
%%%%%%%%%%%%%%%%%%%%%%%%%%%%%%%%%%%

\subsection{Adding electric charge} 
%%%%%%%%%%%%%%%%%%%%%%%%%%%%%%%%%%%

We can generalize the toroidal black hole presented above 
by considering a $U(1)$ gauge field interacting with Einstein gravity. For the action of Maxwell's electrodynamics, we will consider the convention in Ref. \cite{toroidal1}, that is 
\begin{equation}
    I_\text{M}(A_\mu)=-\frac{1}{2\kappa}\int d^{4}x\sqrt{-g}\left(F_{\mu\nu}F^{\mu\nu}\right)\,,\label{Electerm}
\end{equation}
where $F_{\mu\nu}=\partial_{\mu}A_{\nu}-\partial_{\nu}A_{\mu}$, being $A_\mu$ the Maxwell potential. Then, the energy-momentum tensor in Eq. \eqref{Tmunu} has a new contribution, given by 
\begin{equation}
T_{\mu\nu}^{\text{M}}=\frac{2}{\kappa}\left(F_{\mu\rho}F^{\text{ }\rho}_{\nu}-\frac{1}{4}g_{\mu\nu }F_{\rho\sigma}F^{\rho\sigma}\right) \ , 
\end{equation}
and which must be considered in the Einstein equations in Eq. \eqref{Eq2}. We will focus on a Maxwell potential defining a static electric field, that is
\begin{equation}
    A_{\mu}=\left(-\frac{Q}{r},0,0,0\right)\, ,\label{Amu}
\end{equation}
with $Q$ a constant. Using Eq. \eqref{elecpot}, we can show that the charge parameter $Q$ is related to the electrical potential $\Phi$ as follows
\begin{equation}
    \Phi=\frac{Q}{r_{+}}\,.
\end{equation}
where we have used that $\xi^{\mu}$ is the Killing vector $\partial_t$.

It is a direct computation to check that the potential in Eq. \eqref{Amu} is compatible with the static metric in Eq. \eqref{g1}, this means that the Maxwell equations given by
\begin{equation}
    \partial_{\mu}(\sqrt{-g}F^{\mu\nu})=0\,,
\end{equation}
are automatically satisfied, as well as the Skyrme equations in Eq. \eqref{Eq1} (see Ref. \cite{toroidal1}). Then, computing the Einstein equations with the Ansatz in Eqs. \eqref{g1}, \eqref{matter1} and \eqref{Amu}, we lead to the following lapse function
\begin{equation} \label{f2}
f(r)= -\frac{K \kappa q^2 a_N}{4}-\frac{m}{r}  +\frac{(K \kappa q^4 \lambda a_N+32\,Q^{2})}{32 r^{2}}  - \frac{\Lambda}{3} r^2 \ ,
\end{equation}
which solves the complete equations system. This solution generalizes the one presented in Ref. \cite{toroidal1} (see also Refs. \cite{toroidal2} and \cite{toroidal3}) in the context of the $SU(2)$-NLSM. Note that, according to Eq. \eqref{f2}, the Skyrme field contributes in the same way as the Maxwell field, so that one can define an effective coupling constant and then apply all the results derived in Ref. \cite{toroidal4}.

\subsection{Thermodynamics}
%%%%%%%%%%%%%%%%%%%%%%%%%%%%%%%%%%%

The temperature for the charged solution is given by
\begin{equation}\label{Temp2}
    T=-\left(\frac{K\kappa a_{N}q^{2}}{16 \pi r_{+} }+\frac{(K\kappa a_{N} \lambda q^{4}+32 Q^{2})}{128 \pi r_{+}^{3}}-\frac{3 r_{+}}{4 \pi\ell^2}\right) \ ,
\end{equation}
where $r_{+}$ is the radius of the event horizon obtained from Eq. \eqref{f2}, doing $f(r_+)=0$. 

The regularized Euclidean action for the charged solution is given by Eq. \eqref{euact}, but where $I_{\text{bulk}}$ also includes the Euclidean version of the term in Eq. \eqref{Electerm}. After some computations we get to
\begin{equation} \label{Euclact2}
    I^{\text{E}}=-\frac{\beta \pi^2 r_{+}}{4\kappa}\left(K\kappa a_{N}q^{2}+\frac{4 r_{+}^{2}}{\ell^2}-\frac{3K\kappa a_{N}\lambda q^{4}}{8 r_{+}^{2}}+\frac{4 Q^{2}}{r_{+}^{2}}\right)\,.
\end{equation}
It follows that the free energy is given by
\begin{equation} \label{free2}
 G=-\frac{ \pi^2 r_{+}}{4\kappa}\left(K\kappa a_{N}q^{2}+\frac{4 r_{+}^{2}}{\ell^2}-\frac{3K\kappa a_{N}\lambda q^{4}}{8 r_{+}^{2}}+\frac{4   Q^{2}}{r_{+}^{2}}\right)\,.
\end{equation}
Then, we obtain the following thermodynamical potentials
\begin{align}
E&=\frac{2\pi^2}{\kappa}m=-\frac{ \pi^2 r_{+}}{6 \kappa}\left(3K\kappa a_{N}q^{2}-\frac{12 r_{+}^{2}}{\ell^2} -\frac{3K\kappa a_{N}\lambda q^{4}}{8 r_{+}^{2}}-\frac{12 Q^2}{r_{+}^{2}}\right)\,, \label{mass3}\\
    S&=\frac{\pi^{2}r_{+}^{2}}{2}=\frac{A}{4}\,,\\
    \Tilde{Q}&=\frac{\pi Q}{2} \ .
\end{align} 

%%%%%%%%%%%%%%%%%%%%%%%%%%%%%%%%%%%
\section{The charged and rotating toroidal black hole} \label{sec-5}
%%%%%%%%%%%%%%%%%%%%%%%%%%%%%%%%%%%

%%%%%%%%%%%%%%%%%%%%%%%%%%%%%%%%%%%
\subsection{The general solution}

In this section, we will summarize all the results reported in the previous sections to construct the general solution of a charged and rotating toroidal black hole as a solution of the Einstein $SU(N)$-Skyrme model.

The addition of angular momentum to the previously introduced charged solution in Sec. \ref{sec-4} is straightforward. We apply the improper gauge transformation in Eq. \eqref{gaugetrans} to the static black hole metric in Eq. \eqref{g1}, the matter field in Eq. \eqref{matter1}, and to the gauge field in Eq. \eqref{Amu}.
The stationary line element is Eq. \eqref{stationary2}, where the function $f(r)$ is given now by Eq. \eqref{f2}. Then, the  Maxwell potential takes the form
\begin{equation}
    A_{\mu}=\frac{1}{\sqrt{1-\omega^2}}\left(-\frac{Q}{r},0,0,\frac{Q \ell \omega}{r}\right) \,.\label{Amu2}
\end{equation}
The angular velocity is still given by Eq. \eqref{angvel}, while the charge parameter $Q$ is related to the rotational parameter $\omega$ and the electrical potential $\Phi$ in the following way
\begin{equation}
    \Phi=\frac{Q}{r_{+}}\sqrt{1-\omega^{2}}\,.
\end{equation}
Indeed, the electrical potential has the form in Eq. \eqref{elecpot}, but now $\xi^{\mu}$ is the Killing vector $\partial_t+\Omega_{H}\partial_\phi$.

Now we can follow the same steps as in the previous sections to perform the thermal analysis. First, the temperature of the rotating charged solution is
\begin{equation}\label{Temp3}
    T=-\left(\frac{K\kappa a_{N}q^{2}}{16 \pi r_{+} }+\frac{(K\kappa a_{N} \lambda q^{4}+32Q^{2})}{128 \pi r_{+}^{3}}-\frac{3 r_{+}}{4 \pi\ell^2}\right)\sqrt{1-\omega^{2}} \ .
\end{equation}
The regularized Euclidean action is given by Eq. \eqref{Euclact2}, where the period $\beta$ is now depending on the rotational parameter, as can be seen in Eq. \eqref{Temp3}. 
Then, the free energy is given by
\begin{equation} \label{free3}
 G=-\frac{ \pi^2 r_{+}}{4\kappa}\left(K\kappa\alpha_{N}q^{2}+\frac{4 r_{+}^{2}}{\ell^2}-\frac{3K\kappa a_{N}\lambda q^{4}}{8 r_{+}^{2}}+\frac{4 Q^{2}}{r_{+}^{2}}\right)\,.
\end{equation}
 
After some computations, we get that the thermodynamical potential are explicitly given by 
\begin{align}
E&=\frac{\pi^{2}r_+}{32\kappa \left(1-\omega^2\right)}\left[ -8 K\kappa a_N q^{2}(2-\omega^2)+\frac{32 r_{+}^{2}}{\ell^2}\left(2+\omega^2\right)+\frac{ K\kappa a_N \lambda q^{4}+32Q^2}{ 8 r_{+}^2}\left(2-3\omega^2\right)\right]\,,\\ 
J&=-\frac{\ell\pi^2 r_{+}}{4\kappa}\left(K\kappa a_N q^{2}-\frac{12 r_{+}^{2}}{\ell^2}+
\frac{ K\kappa a_N \lambda q^{4}+32 Q^2}{ 8 r_{+}^2}\right)\frac{\omega}{1-\omega^2}\,,\\
S&=\frac{\pi^{2}r_{+}^{2}}{2\sqrt{1-\omega^{2}}}=\frac{A}{4}\,,\\
\Tilde{Q}&=\frac{\pi Q}{2 \sqrt{1-\omega^{2}}} \ ,
\end{align}
ensuring that the first law is satisfied.

\subsection{Global stability in the grand-canonical ensemble}
%%%%%%%%%%%%%%%%%%%%%%%%%%%%%%%%%%%
The thermal stability of a system tells us how their thermodynamic potential behaves under small perturbations from equilibrium.  In the grand-canonical ensemble, the thermodynamic variables are the mass $M$ and the angular momentum $J$, while the temperature $T$ and the angular velocity $\Omega$ are fixed. In this ensemble, the thermodynamical stability analysis can be performed using the Gibbs free energy $G(T,\Omega)$. In order to obtain exact results, here we will consider the case $\lambda=Q=0$, that is, the rotating toroidal black hole as solution of the Einstein $SU(N)$-NLSM. 
We can analyze the global stability by comparing the free energy of the rotating hairy black hole and the vacuum solution with $q=0$, which corresponds to the rotating solution first presented by Lemos in Ref. \cite{Toro1}. Then, we use the grand canonical ensemble by setting both configurations at the same temperature $T$ and angular velocity $\Omega$.

Let us first write the free energy in Eq. \eqref{free} (with $\lambda=0$) as a function of the intensive parameters $T$ and $\Omega$. For this purpose, we invert Eq. \eqref{Temp1} and consider $\omega= \ell\Omega$. Then, we find
\begin{equation}
r_{+}(T)=\frac{4\ell^2\pi T+\sqrt{3K\kappa a_N \ell^2 q^2(1-\ell^2\Omega^2)+16\pi^2\ell^4T^2}}{6\sqrt{1-\ell^2\Omega^2}}\,.  
\end{equation}
It follows that the Gibbs free energy reads
\begin{eqnarray*}
    G(T,\Omega)&=&-\frac{\pi^2}{54\kappa\left(1-\ell^2\Omega^2 \right)^{3/2}}\left(4\pi T \ell^2+\sqrt{16\pi^2 T^2\ell^4+3K\kappa q^2 a_{N}\ell^2(1-\Omega^2  \ell^2 )}\right) \\
  && \left(8\pi^2 T^2 \ell^2 +3K\kappa q^2 a_{N}(1-\Omega^2 \ell^2 )+2\pi \sqrt{16\pi^2 T^4\ell^4+3K\kappa q^2 a_{N}\ell^2T^2(1-\Omega^2  \ell^2 )}\right) .
  \end{eqnarray*}
The above expression is the generalization of the free energy associated to the non-rotating black hole presented in Refs. \cite{toroidal3} \cite{toroidal4}, which can be directly obtained by setting $\Omega=0$. 

\begin{figure}[h!]
\centering
\includegraphics[width=.6\columnwidth]{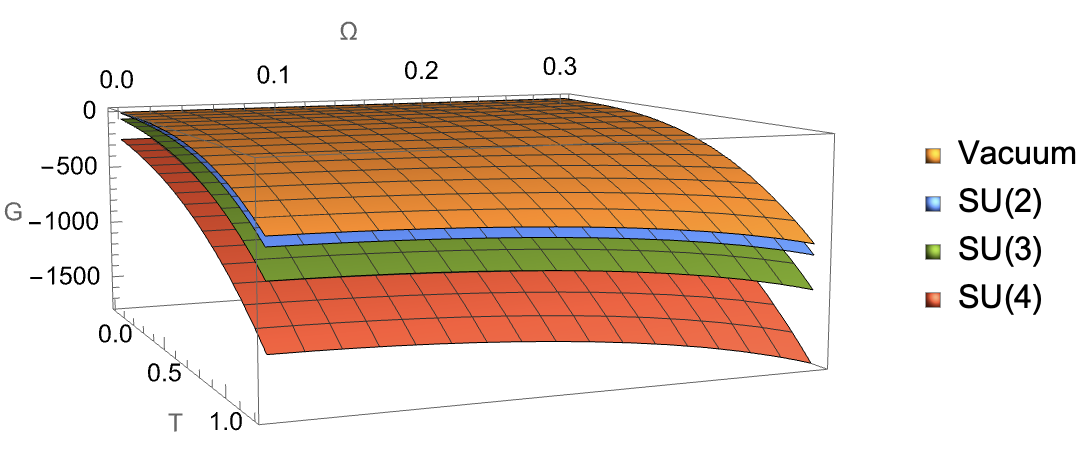} 
\caption{\justifying{Gibbs free energy $G(T,\Omega)$ of the rotating toroidal black hole as a function of the temperature $T$ and angular velocity $\Omega$, for different values of $N$. Here we have fixed the values $K=1; \ell= \sqrt{3}; \kappa= 22$ and $q=1$, except for the yellow surface for which $q=0$. The hairy rotating configuration is always preferred at a given temperature and angular velocity, as its free energy is lower than the one of the vacuum solution.}}
\label{fig1}
\end{figure}
As we can see from Fig. \eqref{fig1}, at the same temperature and angular velocity, the rotating black hole with scalar fields turns out to have a lower value than the undressed black hole. Then, the rotating hairy solution always dominates.\footnote{The analysis of the global stability of the charged non-rotating black hole can be found in Ref. \cite{toroidal1}.}

%%%%%%%%%%%%%%%%%%%%%%%%%%%%%%%%%%%%%
\section{Conclusions} \label{sec-6}
%%%%%%%%%%%%%%%%%%%%%%%%%%%%%%%%%%%%%

In this report, we have shown that the Einstein-Skyrme model in $D=4$ dimensions admits exact solutions describing charged and rotating toroidal black holes. These black holes are asymptotically locally AdS, are characterized by a hair parameter, and allow for an arbitrary number of flavors. The hair of these configurations comes from the Skyrme matter field, in the same way as in the seminal paper of Luckock and Moss in Ref. \cite{Luckock:1986tr}, while the flavor number has been included using the maximal embedding Ansatz of $SU(2)$ into $SU(N)$ in the Euler angles representation. The present solution is the natural generalization of the toroidal black holes constructed in Refs. \cite{toroidal1}-\cite{toroidal4}.  The rotation has been added using an improper gauge transformation. The angular momentum can be calculated in a straightforward manner, and it leads to the fact that the mass of these black holes is bounded from below. On the other hand, the electric charge has been supplemented by considering a $U(1)$ gauge field that interacts with Einstein gravity by a minimal coupling. The electric charge and the Skyrme term go with the same power of $r$ in the lapse function of the black hole, in such a way that it is possible to define an effective coupling constant to perform the thermal analysis. In fact, we have carried out the thermal analysis using the Euclidean approach, showing that not only the geometry depends on the number of flavors considered in the theory, but also the thermodynamic of the present solutions is strongly determined by it. Finally, we have also considered the analysis of the global stability in the grand canonical ensemble finding that the hairy rotating black hole is always preferred over the vacuum solution.

\section*{Acknowlegdments}

The authors are grateful to Andr\'es Anabal\'on and Fabrizio Canfora for many enlightening comments and discussions. This work was funded by the National Agency for Research and Development ANID - SIA grant No. SA77210097 and FONDECYT Grants 11220486, 11241079 and 1231133. ER would also like to thank the Dirección de Investigación and Vice-rectoría de Investigación of the Universidad Católica de la Santísima Concepción, Chile, for their constant support. C. H appreciates the support of FONDECYT postdoctoral Grant 3240632.

\end{document}